# An optical surface resonance may render photonic crystals ineffective

F. García-Santamaría,* Erik C. Nelson, and P. V. Braun

*Department of Materials Science and Engineering, Frederick Seitz Materials Research Laboratory and Beckman Institute,*
*University of Illinois at Urbana-Champaign, Urbana, Illinois 61801, USA*

In this work we identify and study the presence of extremely intense surface resonances that frustrate the coupling of photons into a photonic crystal over crucial energy ranges. The practical utility of photonic crystals demands the capability to exchange photons with the external medium, therefore, it is essential to understand the cause of these surface resonances and a route to their elimination. We demonstrate that by modifying the surface geometry it is possible to tune the optical response or eliminate the resonances to enable full exploitation of the photonic crystal.

## I. INTRODUCTION

Manipulating the flow of light is the overarching application of photonic crystals.[1,2] When describing a particular photonic crystal, the dispersion relation, or band structure, is of paramount importance since it describes the modes to which photons may couple. Despite the tremendous utility, the information contained in the band structure has serious limitations. The band structure only describes the behavior of light within the structure; it does not provide information about the interaction of photons with interfaces between the crystal and the surroundings. This leads to substantial discrepancies between the band structure and the data obtained from external light sources and detectors. For example, due to symmetry considerations, certain modes in the band structure cannot be excited.[3,4] When a photonic crystal is optically probed, it is important to verify that the observed optical features come from the structure and are not the consequence of coupling effects.

In 1902, Wood[5] discovered anomalies in the optical behavior of diffraction gratings that caused abrupt variations of the external observable fields with respect to the wavelength. These anomalies were observed as rapid variations in the intensity of the various diffracted spectral orders in certain frequency bands that could not be explained by ordinary grating theory. In the type of Wood anomalies discussed in this work, the grating surface, surrounded by the lower refractive index substrate and superstrate (air), forms an effective waveguide. If the period of the grating is such that guided modes would be diffracted, the guided modes are no longer bound but become leaky and would escape from the waveguide. If light of a frequency corresponding to one of these leaky guided modes is incident on the surface, a forced resonance occurs resulting in a strong rearrangement of the energy in each diffracted order (see a detailed discussion on the physics of Wood's anomalies in Ref. 6). The spectral positions of the resonances are therefore governed by the geometry of the waveguide and the refractive index of the materials composing it. Recently, Wood's anomalies have been receiving increasing attention for applications in reflectance filters[7] and sensors.[8] 2D and 3D photonic crystals present a surface which is, in essence, a diffraction grating. Therefore, the effect of Wood anomalies is potentially observable in all multidimensional photonic crystals although its intensity will depend on the particular configuration of the system.

In this manuscript, we demonstrate that intense surface resonances can prevent coupling of photons to photonic crystals and render them ineffective. This surface resonance is demonstrated for two different three-dimensional (3D) photonic crystals and is expected to occur in a large number of photonic crystal systems including those made by holography, phase mask lithography, two-photon polymerization in addition to the systems presented here.[9] Such resonances prohibit coupling light to optically active features including resonant cavities or embedded defects[10] within the structure since all photons are reflected at the surface. Likewise, slow photon propagation, negative refraction or super-prism effects would also be negatively affected since they require coupling of photons to specific optical modes.[11] We also report a simple technique to eliminate this resonance and allow the study and application of the underlying band structure.

## II. EXPERIMENTAL

Artificial opals are an example of three-dimensional (3D) photonic crystals that can be engineered to show a photonic band gap (PBG).[12] These PBG structures, known as inverse opals, consist of a face-centered-cubic (fcc) lattice of air spheres in a high dielectric matrix e.g. Si,[13,14] Ge[15] or $Sb_2S_3$.[16] The experimental reflectance spectra from these crystals show strong reflectance peaks in the direction normal to the (111) planes at energies close to those expected for stop bands, one of which is often associated with the existence of a PBG.[13-16] Samples of opal photonic crystals were prepared by vertical deposition[14,17] of 925 nm silica spheres on silicon substrates. This gives a lattice parameter $a$=1.31 μm. A thin layer of aluminum oxide, approximately 8 nm, was grown using atomic layer deposition (ALD) to create the case of interpenetrated spheres.[18] Approximately 86% of the pore volume was loaded with Si by means of chemical vapor deposition (CVD).[19] The optical response of composite opals (SiO$_2$ spheres in Si) and Si inverse opals in the ΓL direction, normal to the (111) planes, is very similar. Since composite opals are more mechanically robust and require less processing, we used them rather than inverse opals for this study.

Reactive ion etching (RIE, Unaxis 790) was performed flowing O2 and SF6 at 20 sccm, with a base pressure of 100 mTorr and applying a field power of 70 W for cycles of 30 s.

The reflectance spectra were taken with a Fourier Transform Infrared (FTIR) system attached to a microscope (Bruker IS66 – Hyperion 2000). A home-made CaF$_2$ 2.4x objective (NA 0.07) was used to illuminate the sample and collect the reflectance from a circular area (diameter ~200 μm). The light source is unpolarized.

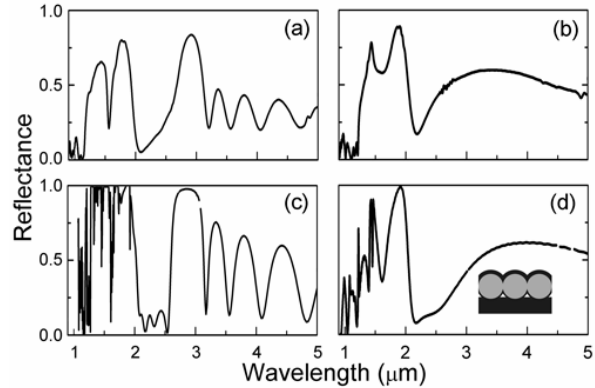

FIG. 1. Reflectance spectra from a SiO$_2$/Si opal collected at regions of a) 9 layers, and b) 1 layer. The FDTD simulations of each case are shown in (c)-(d). The inset in (d) shows a cross-section schematic representation of the model used for the calculations. From the FDTD calculations, the thickness of the overlayer was estimated to be 157 nm.

All 3D finite-difference time-domain (FDTD) simulations were performed with the MIT Electromagnetic Equation Propagation (MEEP) freeware.[20] Calculations account for the Si substrate ($n$=3.5) In the case of artificial opals, the silica ($n$=1.45) sphere radius is 0.36 $a$ to account for the interpenetration (for touching spheres the radius should be $a/\sqrt{8}$). The thickness of the Si film within the opal is set to 0.048 $a$ corresponding to 86% filling of the interstitial sites which is the theoretical maximum that can be achieved through static CVD. Growth of Si on the substrate from the CVD is also taken into account. The thickness of the Si overlayer on top of the first layer of silica spheres can be set independently. The woodpile lattice follows the geometry of the structure demonstrated in Ref 21. This photonic crystal is made of hollow germanium ($n$=4.1) tubes with an internal and external radius of 0.25 $a$ and 0.306 $a$ respectively. The distance between parallel rods is ~0.707 $a$.

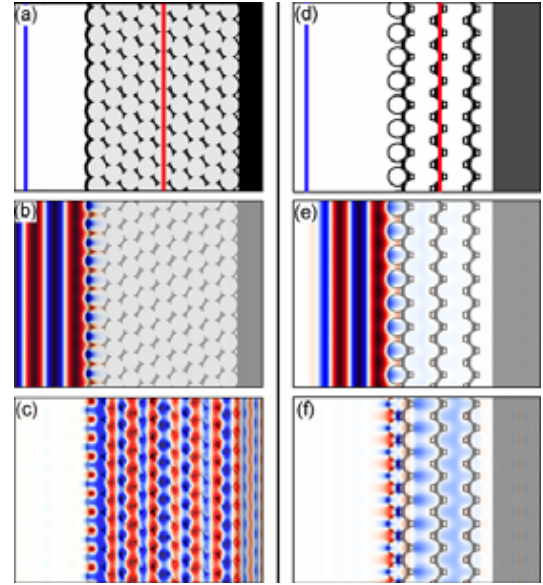

FIG. 2. (Color online) (a) Cross-section of the 3D cell used for the FDTD calculations of a 9 layer SiO$_2$/Si fcc structure with a 0.12 $a$ thick layer of Si on the surface. After 91 periods, the field distribution at $a/\lambda$=0.685 (a maximum of the surface resonance) is simulated for the case of (b) an external light source [blue line in (a)] and (c) and internal source [red line in (a)]. (d-f) show the same as (a-c) for a woodpile structure of hollow Ge tubes with a 0.056 $a$ thick layer of Ge on the surface. The electric field distribution in this case is for a frequency $a/\lambda$=0.812.

### III. RESULTS AND DISCUSSION

#### A. Inability of photonic crystals to exchange photons due to the surface resonance

Experimentally the existence of stop bands in a photonic crystal is usually inferred by analyzing the reflectance spectra.[13-16] Peaks of high reflectance are typically correlated with photonic gaps. Figure 1(a) presents the reflectance spectrum from a 9 layer region of the opal. The spectrum from a monolayer domain of the same sample is shown in Fig. 1(b). As expected, the intensity of the first stop band centered around 2.9 μm shows a strong dependence on the number of layers. However, an intense double peak between 1.4 and 2.0 μm is still present even for the monolayer indicating it originates in the surface of the structure and not in the bulk. In previous works,[13-16] these reflectance peaks were associated with stop bands in the ΓL direction, one of which opens up into a PBG when the silica spheres are removed. If these were in fact stop bands, the intensity of the peaks should diminish as the number of layers is decreased. In reality, these optical features are the result of guided mode resonances, a particular case of Wood's anomalies where the silicon layer over the silica spheres acts as a periodic waveguide. In this instance the resonances reflect nearly all the light from the surface. The reflectance spectra agree well with 3D finite-difference time-domain (FDTD) simulations [Fig. 1 (c-d)]. At normal incidence, polarization does not have a significant effect on the intensity of the reflectance, consistent with other studies on 2D gratings.[22]

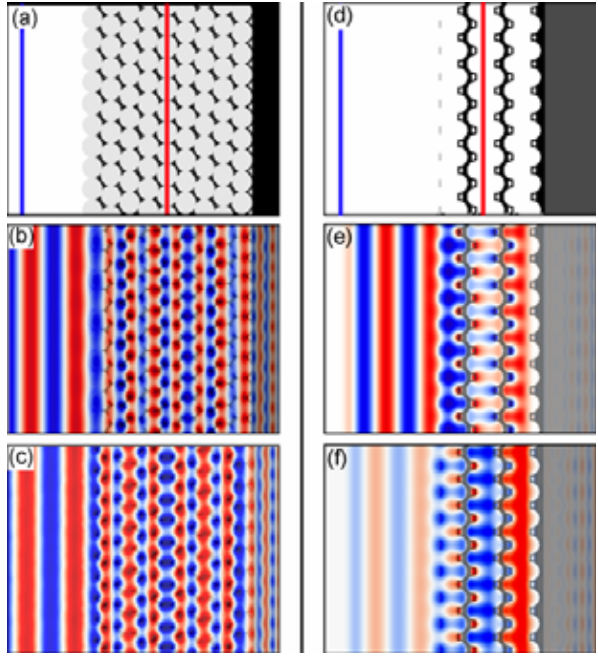

FIG. 3. (Color online) (a) Cross-section of the 3D cell used for the FDTD calculations of a 9 layer $SiO_2$/Si fcc structure where the overlayer has been completely removed. After 91 periods, the field distribution at $a/\lambda=0.685$ (a maximum of the surface resonance) is simulated for the case of (b) an external light source [blue line in (a)] and (c) and internal source [red line in (a)]. (d-f) show the same as (a-c) for a woodpile structure of hollow Ge tubes with a 0.056 $a$ thick layer of Ge on the surface. The electric field distribution in this case is for a frequency $a/\lambda=0.812$.

To further understand the nature of this resonance, the electric field distribution was plotted for two different photonic crystals at a frequency where the resonance reflects more than 98% of the radiation. Figure 2(a) shows the cross-section of the 3D cell used for the case of a $Si/SiO_2$ opal with a thick layer (0.12 $a$) of Si on the surface. The silicon substrate is also taken into account. In this case the light frequency is $a/\lambda=0.685$. For an external monochromatic light source, blue line in Fig. 2(a), the electric field resonates within the Si overlayer and is reflected; almost no photons couple into the structure [Fig. 2(b)]. Likewise, for an internal[23] monochromatic light source, red line in Fig. 2(a), the electric field undergoes a similar resonance and cannot couple to the external medium [Fig. 2(c)]. As previously mentioned, the surface resonances due to Wood's anomalies are expected to be observed in any photonic crystal whose surface presents both a grating and a waveguiding structure (high index material grown on a lower index template). An inverse woodpile structure[24] consisting of hollow tubes of a high dielectric constant material is an especially interesting example since it is expected to present a very large PBG (over 25% for Si and 30% for Ge).[25] Here we present the case of hollow germanium (n=4.1) tubes with a 0.056 $a$ Ge overlayer.[19] The field distribution at $a/\lambda=0.812$ for both an internal and an external light source is shown in Figures 2(d-f) together with a cross section of the 3D calculations cell. Again, the resonance reflects back almost all the radiation for both an internal and an external light source. This demonstrates that the resonance is observable for photonic crystals with different geometries and further highlights the dramatic inhibition of photons exchange between the photonic crystal and the external medium.

### B. Elimination of the surface resonance

Since the surface resonances investigated here are a result of leaky guided modes in the high refractive index overlayer, it is reasonable to expect a substantial change in the resonance intensity if the overlayer is removed. Figure 3(a) shows the cross-section of the 3D cell used for the case of a $Si/SiO_2$ opal where the Si overlayer has been removed. The fields in Figs. 3(b-c) propagate now from the external medium to the artificial opal and vice-versa. The same effect is observed in an inverse Ge woodpile structure where the top half of the first layer of cylinders has been removed [Fig. 3(d-f)].

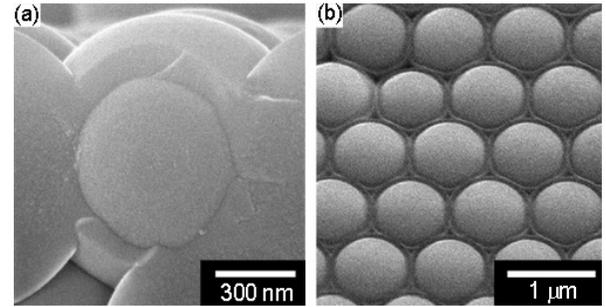

FIG. 4. (a) SEM image of a cross-section of the first layer of a $Si/SiO_2$ opal. (b) SEM image showing the surface of the opal after performing RIE to remove the Si overlayer.

Experimentally, a directional and controlled etch of the high refractive index overlayer can be achieved by means of reactive ion etching (RIE).[14] Figure 4(a) shows a cross-section of the first layer of a $Si/SiO_2$ opal where the Si overlayer can be clearly observed. With RIE, the overlayer is etched and the top half of the $SiO_2$ spheres exposed [Fig. 4(b)].

Once the surface resonance has been significantly attenuated it becomes possible to probe the band structure, couple photons to the available modes and experimentally detect the presence of stop bands. In Fig. 5, the reflectance spectra of a 9 layer $Si/SiO_2$ opal have been simulated by FDTD for (a) the case of a sample with a 157 nm layer of Si on the surface and (b) the case of exposed $SiO_2$ hemispheres. The shaded areas correspond to the stop bands calculated from the band structure [Fig. 5(e)].[26] It is observed that for the case of exposed $SiO_2$ hemispheres, the reflectance peaks match extraordinarily well with the stop bands. For wavelengths out of the stop bands, low reflectances or Fabry-Perot oscillations which denote that photons are indeed coupling to the system,[3] are observed. Figures 5 (c, d) present the experimentally measured reflectance spectra from a 14 layer sample before and after removing the Si overlayer by RIE. As predicted by the FDTD calculations, the resonance intensity decreases once the Si overlayer is removed and the reflectance peaks due to the stop bands are visible in the expected spectral positions. The Fabry-Perot oscillations between 1.3 and 1.5 μm in Fig. 5(b) are not observed experimentally because these photons are diffracted in directions that our finite numerical aperture objective does not collect.[27] The intensities of the reflectance peaks for the etched sample [Fig. 5(d)] are appreciably lower than those predicted by simulations. This is likely due to diffuse scattering from defects and disorder within the original artificial opal,[28,29] indicating that the quality of the artificial opals may need to be improved for demanding applications.

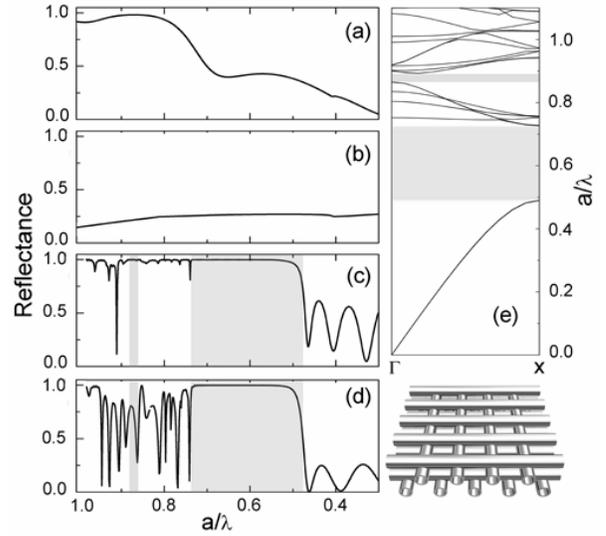

FIG. 6. FDTD reflectance spectra calculated for one layer of a woodpile structure of hollow Ge rods. (a) 0.056a thick film of Ge on the surface, and (b) top half of the Ge tubes is removed. 12 layers of the same structure are shown in for the case of (c) 0.056a thick film of Ge on the surface, and (d) top half of the Ge first layer of tubes is removed. The shaded areas represent the stop bands in the ΓX direction calculated from the band structure (e). The inset shows a rendering of an inverse woodpile structure.

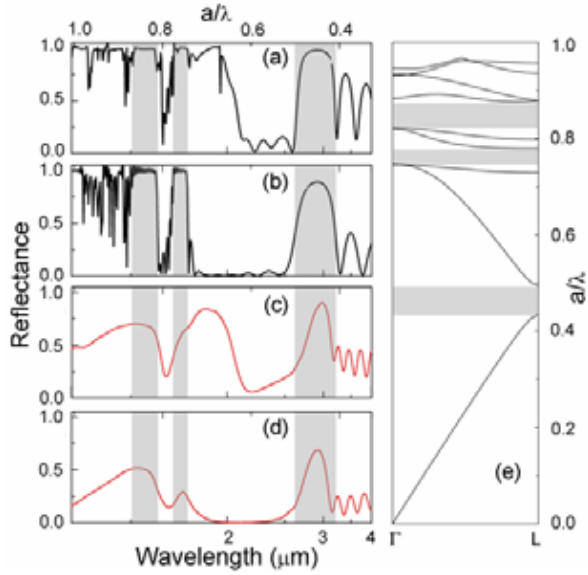

FIG. 5. (Color online) FDTD calculated reflectance spectra from a 9 layers Si/SiO$_2$ opal with a (a) 157 nm thick film of Si on the surface, and (b) exposed SiO$_2$ hemispheres. Experimental reflectance spectra from a 14 layers Si/SiO$_2$ opal (c) after Si CVD, and (d) after a 120 second cycle of RIE. The shadowed areas represent the stop bands in the ΓL direction calculated from the band structure (e).

As shown before, the inverse Ge woodpile is also affected by the presence of a surface resonance. Figure 6(a) is the FDTD simulated reflectance spectrum from a single layer of this structure (electric field parallel to the tubes). Nearly 100% reflectance is obtained over a large frequency range. When the Ge overlayer is removed, the reflectance is reduced significantly [Fig. 6(b)]. The reflectance spectra are also simulated for a 12 layer sample before and after removing the Ge overlayer [Fig. 6 (c, d)]. When the Ge overlayer is present, the range of frequencies for which reflectance is close to unity is significantly larger than the stop band predicted by band structure calculations [Fig. 6(e)]. In similar fashion to Si/SiO$_2$ artificial opals, once the overlayer is removed [Fig. 6 (d)], the reflectance peak matches the fundamental gap and photons of higher energies couple to available modes.

## C. Effects of the overlayer thickness on the optical behavior of the resonance

The thickness and refractive index of the layer deposited on top of the structure determine to a great extent the intensity, spectral width and wavelength of the observed surface resonance.[7,8] In Fig. 7(a-e) the reflectance spectra of a monolayer of silica spheres with a 0.12 $a$ thick Si overlayer are shown after an increasing number of RIE cycles. The calculated spectra are plotted in Fig. 7(f-j). When the Si overlayer is present, the resonances are strong and their shape is relatively invariant. Removing part of the Si only causes a relatively minor blue-shift of the spectral position. As the silica spheres are exposed, the shape and intensity of the peaks show more obvious changes. Once the overlayer is completely etched, resonant guided modes are no longer available and the resonance is dramatically attenuated; photons are transmitted to the underlying system.

The thickness of the high refractive index overlayer can be adjusted to tune the resonance wavelength and, more importantly, its intensity. In Fig. 8 the calculated frequency and reflectance intensity of the lowest energy resonance (the lower energy peak of the double peaked resonance) for a spheres monolayer is plotted as a function of the overlayer thickness. The calculations are performed for different refractive indices of both the overlayer and the spheres. Several conclusions can be extracted from these graphs. The intensity and spectral position of the resonance does not depend on the spheres refractive index once the overlayer is thick enough. For example, in the case of our artificial opals with a 157 nm (0.12 $a$) thick overlayer, the resonance will look almost identical if the silica was etched to obtain an inverse opal. Interestingly, the presence of very thin films (a few nanometers) of Si on the surface of the opal is sufficient to yield extremely intense resonances. In addition, increasing the overlayer thickness red-shifts the spectral position of the reflectance peaks. Ref. 19 illustrates how the intensity of the resonance (incorrectly assigned to a stop band) increases and red-shifts as high refractive index materials are grown. Also, it is interesting to note that, in order to fit the experimental data in Fig. 1 (a and b) the theoretical model required the film thickness to be 157 nm, approximately 2.5 times thicker than the Si layer within the opal (63 nm). This is reasonable since CVD is a deposition technique that enables nearly homogeneous growth of Si within the artificial opal, but often displays a higher growth rate at the surface. Also, once the internal pores are closed, Si keeps growing on the surface of the structure until the CVD is terminated. It can be seen that the intensity of the resonance is slightly lower than the values predicted by the simulations, most likely due to diffuse scattering. However, the intensities are comparable to or greater than those obtained by systems that were specifically designed to show this resonance.[7,22] The data shown in Fig. 8(c,d) for an overlayer with $n$=2.6 indicate that polymer structures with a film of $TiO_2$ should show these resonances unless the oxide film is removed as in ref. 30.

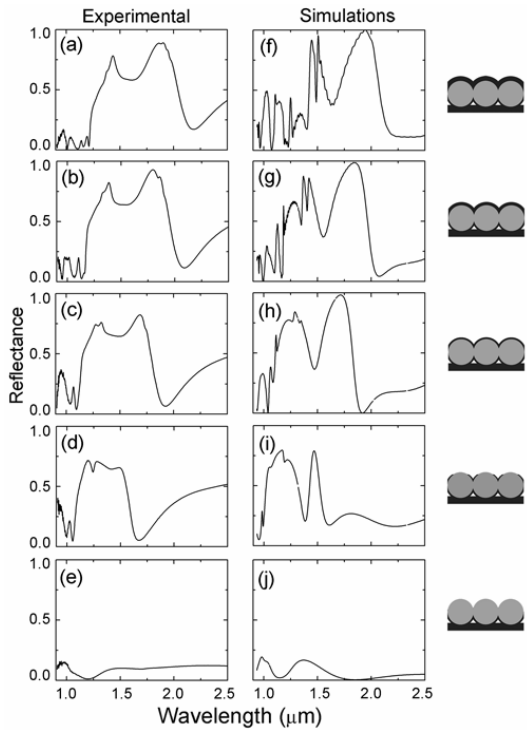

FIG. 7. Behavior of the resonance as a function of the overlayer thickness. Comparison of reflectance spectra from a monolayer of Si/SiO2 spheres a) as grown and after b) 1, c) 2, d) 4 and e) 5 cycles of RIE (30 seconds each). FDTD simulations are provided for varying silicon overlayer thicknesses on a monolayer of close-packed spheres. The radius of the overlayer is 0.48 $a$ and the center is offset with respect to the center of the silica spheres by 0.0 $a$, 0.02 $a$, 0.05 $a$, 0.10 $a$ and 0.48 $a$ from (f) to (g) respectively. The insets on the right show a cross-section of the cells used for the FDTD simulations.

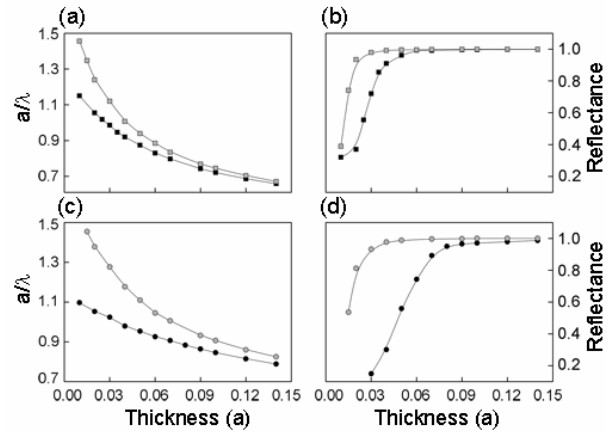

FIG. 8. FDTD calculations for the (a) spectral position (expressed as normalized frequency) and (b) reflectance intensity of the lowest energy resonance as a function of the Si overlayer thickness for a monolayer of silica (solid squares) and air (hollow squares) spheres. (c) and (d) show the same relationships for a monolayer of polystyrene ($n$=1.6, solid circles) and air (hollow circles) spheres with a $TiO_2$ ($n$=2.6) overlayer. Notice the lattice parameter is that of an fcc structure and it is equal to √2 times the distance between neighbor spheres. Lines are drawn in each plot as a guide to the eye.

## IV. CONCLUSIONS

In summary, we have demonstrated the existence of a surface resonance due to Wood's anomalies that renders photonic crystals ineffective as it prevents the exchange of photons between the crystal and an external medium. This resonance leaves the modes of the band structure inaccessible for broad frequency ranges of great optical interest. A very convenient method to suppress the resonance by etching the surface has been demonstrated in two different structures. In the particular case of artificial opals, eliminating the resonance allows direct measurement of two high energy stop gaps that are otherwise inaccessible. The effect of Wood's anomalies is very general and potentially observable in many photonic crystals with different symmetries as long as the geometry of the surface is capable of acting as a waveguide (particularly favorable is the case of inverted geometries). It is critical to acknowledge the presence of the resonance and eliminate it. Phenomena that rely on coupling photons to particular bands, including embedded cavities, negative refraction and the super-prism effect, will not be achievable unless the resonance is attenuated.

## V. ACNOWLEDGMENTS

This material is based in part on work supported by the ARO-MURI grant DAAD19-03-1-0227, and the DOE, through the FS-MRL at UIUC grant DEFG02-91ER45439.
* Electronic address: floren@uiuc.edu.